\documentclass{article}
\usepackage[dvips]{graphicx}

\begin{document}

\noindent
{\LARGE
Trapped, Two-Armed, Nearly Vertical Oscillations in Polytropic Disks
}
\bigskip
\begin{center}
Shoji \textsc{Kato}

2-2-2 Shikanodai, Ikoma-shi, Nara, Japan 630-0114

email: kato.shoji@gmail.com

(to be published in PASJ 62, No.3, 2010)
\end{center}


%



\begin{abstract}

We have examined trapping of two-armed nearly vertical oscillations in polytropic disks.
Two-armed nearly vertical oscillations are interesting in the sense that they are 
trapped in an inner region of disks with proper frequencies, if the inner edge of disks 
is a boundary that reflects oscillations.
The frequencies of the trapped oscillations cover the frequency range of kHz QPOs to 
low frequency QPOs in LMXBs, depending on the modes of oscillations.
Low frequency trapped oscillations are particularly interesting since their trapped region is wide.
These low frequency oscillations are, however, present only when $\Gamma(\equiv 1+1/N)$ is close to 
but smaller than 4/3 (when spin parameter $a_*$ is zero), where $N$ is the polytropic index. 
The above critical value 4/3 slightly increases as $a_*$ increases.  

\end{abstract}

\section{Introduction}

One of the possible candidates of high-frequency quasi-periodic
oscillations (HFQPOs) observed in low-mass X-ray binaries is resonantly excited p- and/or g-mode oscillations
in deformed disks (Kato 2004, 2008a,b; Ferreira and Ogilvie 2008; Oktariani et al. 2010).
The disk deformation required in this model is a warp or an eccentric deformation in equatorial plane.
Recent numerical MHD simulations (Henisey et al. 2009) seem to have preliminary 
confirmed the presence of the excitation mechanism.

The high frequency QPOs observed in black-hole candidates appear in pairs with frequency ratio of 3 : 2.
This characteristic of high-frequency QPOs in black-hole candidates seems to be described by the above model,
if we assume that geometrically thin disks are surrounded by hot tori and that 
the QPOs photons are disk photons Comptonized in the tori (Kato and Fukue 2006).
The twin kHz QPOs observed in neutron stars, on the other hand, have always not the 3 : 2 frequency ratio.
They change their frequencies with time with correlation. 
If we want to describe such characteristics of kHz QPOs by the above model, 
a warp or eccentric disk deformation must have a time-dependent precession.
In the case of neutron stars, distinct from the case of black holes, such precession of the deformation 
might be expected, since the central stars have surfaces and this might become causes of time-dependent precession 
through magnetic and radiative couplings between the central sources and the disks.
It is not clear, however, whether possible time-dependent precession of deformation has a time scale consistent 
with the time variations of the kHz QPOs.

In this context, it is worthwhile examining whether there are other kinds of disk deformations 
that can become a possible source of resonant excitation of high-frequency oscillations in disks.
This problem has been examined by Kato (2009), and it was suggested that two-armed nearly vertical
oscillations\footnote{
   There will be no commonly accepted classification and terminology concerning disk oscillations.
   Here, we classify oscillations into  four types, i.e., p-mode, g-mode, c-mode and 
   vertical p-mode oscillations (see, for example, Kato 2001; Kato et al. 2008).
   The oscillations considered here are the vertical p-mode oscillations. } 
can excite high-frequency p- and/or g-mode oscillations.
Hence, it will be interesting to examine whether there are two-armed nearly vertical oscillation modes that
can become a global deformation of disks. 

More interestingly, the two-armed vertical disk oscillations themselves may be origins of the varous 
types of QPOs observed in neutron-star LMXBs.
This is because two-armed nearly vertical osillations occur in the inner region of disks and 
cover a wide range of frequency by difference of modes.

Based on these considerations, we examine in this paper 
basic properties of the two-armed vertical disk oscillations.
That is, we examine their characteristics of trapping, eigen-frequency and its dependence on polytropic index 
specifying the vertical disk structure, using the mathematical formulations already prepared by 
Silbergleit et al. (2001) to study the trapping of the corrugation waves (c-mode oscillations).

\section{Overview of Trapping of Two-Armed Vertical Oscillations in Polytropic Disks}

Before having a rough image of nearly vertical oscillations in vertically polytropic disks, we briefly 
mention nearly vertical oscillations in isothermal disks, because it is instructive to have a 
rough image of wave trapping in disks.

If a disk oscillation with azimuthal wavenumber $m$ occur isothermally and locally in vertical direction in 
isothermal disks, the frequency, $\omega$, is given by (see, e.g., Okazaki et al. 1987)
\begin{equation}
    (\omega-m\Omega)^2-n\Omega_\bot^2=0,
\label{1.1}
\end{equation}
where $n$ denotes the node number ($n=1,2,...$) of the oscillations in the vertical direction, 
and $\Omega$ and $\Omega_\bot$ are, respectively, the angular 
velocity of disk rotation at the radius in consideration and the vertical epicyclic frequency at the radius.
The oscillations are, however, cannot be purely vertical by inhomogeneity of the disk in the radial direction.
They essentially have horizontal velocity components. 
In other words, the vertical oscillations described by equation (\ref{1.1}) are not rigorous.
The vertical oscillations couple with purely horizontal inertial oscillations, 
$(\omega-m\Omega)^2-\kappa^2=0$, through pressure.
If the coupling is considered with local approximations in the radial direction, we have a dispersion relation,
which is written as (Okazaki et al. 1987)
\begin{equation}
  [(\omega-m\Omega)^2-\kappa^2][(\omega-m\Omega)^2-n\Omega_\bot^2]=c_{\rm s}^2k^2(\omega-m\Omega)^2,
\label{1.2}
\end{equation}
where $c_{\rm s}$ is the acoustic speed, $\kappa$ is the horizontal epicyclic frequency, and
$k$ is the horizontal wavenumber of the oscillations.

Since we are now interested in nearly vertical oscillations, i.e., $(\omega-m\Omega)^2-n\Omega_\bot^2\sim 0$
($n=$ 1, 2, 3,...), equation (\ref{1.2}) shows that their propagation region in the radial direction 
is described by
\begin{equation}
     (\omega-m\Omega)^2-n\Omega_\bot^2\geq 0,
\label{1.3}
\end{equation}
since $\Omega_\bot$ is always larger than $\kappa$.

We now proceed to polytropic disks with barotropic gas with polytropic index $N$, 
i.e., $p\propto \rho^{(1+1/N)}$, where $p$ and $\rho$ are pressure and density, respectively.
The frequency, $\omega$, of purely vertical oscillations is then given by (Silbergleit et al 2001, Kato 2005)
\begin{equation}
       (\omega-m\Omega)^2-\Psi_n\Omega_\bot^2=0,
\label{1.4}
\end{equation}
where $\Psi_n$ is 
\begin{eqnarray}
    \Psi_n=\left\{\begin{array}{ll} 1 & {\rm for}\quad n=1\\
                                     2+1/N & {\rm for}\quad n=2 \\
                                     3+3/N & {\rm for}\quad n=3.        \label{1.5}    
                    \end{array}
      \right. 
\end{eqnarray}     
The case of isothermal disks corresponds to $N=\infty$.

In oscillations of $n=1$, $\omega-m\Omega$ is independent of $N$, but in oscillations of $n\not= 1$
it depends on $N$.
That is , for two-armed oscillations ($m=2$), equation (\ref{1.4}) has solution given by 
\begin{equation}
     \omega=2\Omega-(1+\Gamma)^{1/2}\Omega_\bot \qquad {\rm for} \quad n=2
\label{1.6}
\end{equation}
and
\begin{equation}
    \omega=2\Omega-(3\Gamma)^{1/2}\Omega_\bot \qquad {\rm for} \quad n=3,
\label{1.7}
\end{equation}
where $\Gamma=1+1/N$, and $\Gamma$ is usually in the range of $\Gamma=1\sim 5/3$.
An interesting result shown in equation (\ref{1.7}) is that oscillations with $n=3$ can have low-frequencies 
for reasonable values of $\Gamma$.

As in the case of isothermal disks, the nearly vertical oscillations are also not localized at a particular 
radius.
They propagate in the radial direction.
The propagation region is specified, as in the case of isothermal disks, by (see section 4 for details)
\begin{equation}
    (\omega-m\Omega)^2 - \Psi_n\Omega_\bot^2\geq 0.
\label{1.8}
\end{equation}
The propagation diagram showing the propagation region of the oscillations in the radial direction is given 
in figure 1 for oscillations with $n=1$ and 2, and in figure 2 for oscillations with $n=3$.
As the value of $n$ increases, the propagation region of oscillations becomes narrower in the 
frequency - radius plane (i.e., propagation diagram), and finally for oscillations 
with $n=3$, the propagation region in the inner region of disks disappears for oscillations with $\Gamma> 4/3$.

The next problem is to examine where the oscillations are really trapped and how much the frequency of the trapped 
oscillations is.
To examine this problem, a careful treatment of the wave equation is necessary 
and this is done in the subsequent sections.
Here, however, in order to have a rough image of trapping, we discuss qualitatively what determines 
the frequency of the trapped oscillations.

Let us assume tentatively that the frequency of the trapped oscillation is $\omega$ and the outer boundary 
of the trapped region is $r_{\rm c}$.
The inner boundary of the trapped region, $r_{\rm i}$, is taken at the inner edge of the disks, 
which is $3r_{\rm g}$ when the central source has no spin,
$r_{\rm g}$ being the Schwarzschild radius.
Then, roughly speaking, $r_{\rm c}-r_{\rm i}$ must be on the order of the wavelength of the
trapped oscillations.
If some kinds of mean values of $(\omega-m\Omega)^2-\kappa^2$ and $c_{\rm s}^2$ in the trapped region
are denoted, respectively, $\langle (\omega-m\Omega)^2-\kappa^2\rangle$ and $\langle c_{\rm s}^2\rangle$,
then the trapping condition is $\langle(\omega-m\Omega)^2-\kappa^2\rangle \sim \langle c_{\rm s}^2\rangle/
(r_{\rm c}-r_{\rm i})^2$ (for the fundamental mode in the radial direction), 
since oscillations propagate in the radial direction as inertial acoustic waves.
This trapping condition is realized only when $\omega$ has a particular value, say $\omega_{\rm t}$.
This is because for $\omega>\omega_{\rm t}$, 
$\langle (\omega-m\Omega)^2-\kappa^2\rangle(r_{\rm c}-r_{\rm i})^2$ is too small compared with 
$\langle c_{\rm s}^2\rangle$,
since the width of the propagation region, i.e., $r_{\rm c}-r_{\rm i}$, is too narrow (see figures 1 and 2).
On the other hand, for $\omega<\omega_{\rm t}$, the value of 
$\langle (\omega-m\Omega)^2-\kappa^2\rangle(r_{\rm c}-r_{\rm i})^2$ becomes larger than 
$\langle c_{\rm s}^2\rangle$, since the width, $r_{\rm c}-r_{\rm i}$, becomes too large.
To know the detailed value of $\omega_{\rm t}$ for the trapping, we must make numerical calculations,
as is done in the subsequent sections.

At this stage, without detailed numerical calculations, we can say that when $\Gamma$ is large, 
the frequency of the trapped oscillations is low, if other parameters, including the radial distribution 
of $c_{{\rm s}0}$ in disks, are fixed (see figure 4).
Let us assume that the trapping condition has satisfied for a frequency, say $\omega_{\rm t}$, 
when $\Gamma$ has a value, say $\Gamma_{\rm t}$.
If $\Gamma$ increases from $\Gamma_{\rm t}$, $r_{\rm c}$ and thus $r_{\rm c}-r_{\rm i}$ decreases 
from the value in the above case, if the frequency $\omega$ is fixed at $\omega_{\rm t}$
[see equations (\ref{1.6}) and (\ref{1.7}) and curves in figures 1 and 2].
Thus, the trapping condition is not satisfied at $\omega=\omega_{\rm t}$.
For the trapping condition to be satisfied we must increase the width of the trapping region.
This can be done by decreasing $\omega$ from $\omega_{\rm t}$.

\section{Unperturbed Disks and Equations Describing Disk Oscillations}

\subsection{Basic Assumptions and Unperturbed Disks}

We consider geometrically thin, relativistic disks.
For mathematical simplicity, however, the effects of general relativity are taken into 
account only when we consider radial distributions of $\Omega(r)$, $\kappa(r)$, and $\Omega_\bot(r)$. 
They are, in turn, the angular velocity of disk rotation, the epicyclic frequencies in the radial and 
vertical directions. 
Except for them, the Newtonial formulations are adopted.
Since geometrically thin disks are considered, $\Omega$, is approximated to be the relativistic 
Keplerian angular velocity, $\Omega_{\rm K}(r)$, when its numerical values are necessary.
Here, $r$ is the radial coordinate of cylindrical ones ($r$,$\varphi$,$z$), where the $z$-axis is
perpendicular to the disk plane and its origin is the disk center.
Functional forms of $\Omega_{\rm K}(r)$, $\kappa(r)$, and $\Omega_\bot(r)$ are given 
in many literature sources (e.g., Kato et al. 2008)

The disks are assumed to consist of a barotropic gas with polytropic index $N$, i.e., 
the pressure $p$ and density $\rho$ are related by $p\propto \rho^{(N+1)/N}$.
The hydrostatic balance in the vertical direction then gives (e.g., Kato et al. 2008)
\begin{equation}
   \rho_0(r,z)=\rho_{00}(r)\biggr(1-\frac{z^2}{H^2}\biggr)^N,
\label{2.1}
\end{equation}
\begin{equation}
   p_0(r,z)=p_{00}(r)\biggr(1-\frac{z^2}{H^2}\biggr)^{1+N},
\label{2.2}
\end{equation}
where subscript 0 represents the quantities in the equilibrium state and 00 are those on the 
equatorial plane.
The acoustic speed defined by $c_{\rm s}^2=\Gamma dp_0/d\rho_0$ is also given by
\begin{equation}
   c_{\rm s}^2(r,z)=c_{\rm s0}^2(r)\biggr(1-\frac{z^2}{H^2}\biggr),
\label{2.3}
\end{equation}
where $\Gamma$ is related to $N$ by $\Gamma=(N+1)/N$ or $N=1/(\Gamma-1)$.
The hydrostatic balance in the vertical direction also shows that the half-thickness $H$ 
of disks is related to $c_{{\rm s}0}$ and $\Omega_\bot$ by
\begin{equation}
       \Omega_\bot^2H^2=2Nc_{{\rm s}0}^2.
\label{2.3a}
\end{equation}

\subsection{Equations Describing Disk Oscillations}

We consider small-amplitude perturbations in the above disk.
The perturbations are assumed to be proportional to exp$[i(\omega t-m\varphi)]$, where $\omega$ and $m$ are 
frequency and azimuthal wavenumber, respectively, of the perturbations.
Then, hydrodynamical equations describing them are (e.g., Kato et al. 2008)
\begin{equation}
    i{\tilde \omega}\rho_1+\frac{\partial}{r\partial r}(r\rho_0u_r)-i\frac{m}{r}\rho_0u_\varphi
        +\frac{\partial}{\partial z}(\rho_0u_z)=0,
\label{3.1}
\end{equation}
\begin{equation}
      i{\tilde \omega}u_r-2\Omega u_\varphi =-\frac{\partial h_1}{\partial r},
\label{3.2}
\end{equation}
\begin{equation}
     i{\tilde \omega}u_\varphi+\frac{\kappa^2}{2\Omega}u_r=i\frac{m}{r}h_1,
\label{3.3}
\end{equation}
\begin{equation}
     i{\tilde \omega}u_z=-\frac{\partial h_1}{\partial z},
\label{3.4}
\end{equation}
where $h_1$ and ${\tilde \omega}$ are defined, respectively, by 
\begin{equation}
      h_1=\frac{p_1}{\rho_0}=c_{\rm s}\frac{\rho_1}{\rho},
\label{3.5}
\end{equation}
\begin{equation}
     {\tilde \omega}=\omega-m\Omega.
\label{3.6}
\end{equation}
Here, $(u_r,u_\varphi,u_z)$, $p_1$, and $\rho_1$ are 
the Eulerian velocity, pressure, and density perturbations over the unperturbed ones,
respectively.
Equation (\ref{3.1}) is the equation of continuity, 
equations (\ref{3.2}) -- (\ref{3.4}) are in turn the $r$-, $\varphi$-, and $z$- 
components of equation of motion.

Elimination of $u_r$, $u_\varphi$, $u_z$, and $\rho_1$ from equations
(\ref{3.1}) -- (\ref{3.5}) leads to a partial differential equation of $h_1$.
In the processes, we neglect the radial variations of the unperturbed density and some other quantities, 
assuming that their characteristic radial scales are longer than 
the characteristic radial wavelength of the perturbations.
Then, we have an equation describing the variation of $h_1$ in the form:
\begin{equation}
   \frac{1}{{\tilde \omega}}\frac{\partial}{\partial r}\biggr(\frac{{\tilde \omega}}{{\tilde \omega}^2
      -\kappa^2}\frac{\partial h_1}{\partial r}\biggr)
      +\frac{1}{{\tilde \omega}^2}\frac{1}{\rho_0}\frac{\partial}{\partial \rho_0}\frac{\partial}{\partial z}
          \biggr(\rho_0\frac{\partial h_1}{\partial z}\biggr)
      +\frac{1}{c_s^2}h_1 =0.
\label{3.7-}
\end{equation}
As is clear from the derivation processes, this equation is obtained under the framework of the Newtonian hydrodynamics.
A fully general relativistic one corresponding to equation (\ref{3.7-}) has been derived by Perez et al.
(1997) and used by Silbergleit et al. (2001, 2008).
In their equation the factor ${\tilde \omega}$ inside and the factor $1/{\tilde \omega}$ outside the
first $r$-derivative $\partial/\partial r$ in the first term on the right-hand side of equation (\ref{3.7-})
are absent.
However, relativistic factors related to metric appear in the first and third terms of equation (\ref{3.7-}).
These factors are on the order of unity, but not unity.
For example, on the first term of equation (\ref{3.7-}), a metric factor $g^{rr}(r)$ is multiplied, 
which is $2/3$ at $3r_{\rm g}$ 
($r_{\rm g}$ being the Schwarzschild radius) in the case of the Schwarschild metric.
   
In this paper we do not take into account the relativistic effects except the functional forms of
$\Omega$, $\kappa$ and $\omega_\bot$ as mentioned before and start from equation (\ref{3.7-}).
In the case of the polytropic gas mentioned in subsection 3.1, by changing 
the independent variables from ($r$, $z$) to ($r$, $\eta$),
where $\eta=z/H$, we can reduce equation (\ref{3.7-}) to 
\begin{equation}
    {\tilde \omega}H^2\frac{\partial}{\partial r}\biggr(\frac{{\tilde \omega}}{{\tilde \omega}^2-\kappa^2}
         \frac{\partial h_1}{\partial r}\biggr)
         +\frac{1}{(1-\eta^2)^N}\frac{\partial}{\partial\eta}\biggr[(1-\eta^2)^N
         \frac{\partial h_1}{\partial \eta}\biggr] 
         +\frac{{\tilde \omega}^2H^2}{c_{\rm s0}^2}\frac{1}{1-\eta^2}h_1=0.
\label{3.7}
\end{equation}

The assumption of strong variation of perturbations in the radial direction ensures the 
WKB separability of variable in the above partial differential equation in the form of
$h_1(r,\eta)={\tilde h}_1(r)g(r,\eta)$.
Here, the function $g$ varies slowly with $r$.
Extensive studies of disk oscillations by the 
WKB procedures have been made by Wagoner and his collaborators, 
e.g., Nowak and Wagoner (1992), Perez et al. (1997),
Silbergleit et al. (2001).
Here, we follow the procedures of Silbergleit et al. (2001) used to study trapped $c$-mode 
oscillations, although notations adopted here somewhat different from theirs.

By using the decomposition of $h_1(r,z)$ mentioned above, we can reduce equation (\ref{3.7}) into the 
following set of two ordinary differential equations (Silbergleit et al. 2001):
\begin{equation}
    \frac{1}{(1-\eta^2)^N}\frac{d}{d\eta}\biggr[(1-\eta^2)^N\frac{dg}{d\eta}\biggr]
      +\frac{{\tilde \omega}^2H^2}{c_{\rm s0}^2}\frac{1}{1-\eta^2}g-2NKg=0,
\label{3.8}
\end{equation}
and
\begin{equation}
     {\tilde \omega}H^2\frac{d}{dr}\biggr[\frac{{\tilde \omega}}{{\tilde \omega}^2-\kappa^2}
        \frac{d{\tilde h}_1}{dr}\biggr] +2NK{\tilde h}_1=0,
\label{3.9}
\end{equation}
where $2NK$ is the separation constant.
Since the term ${\tilde \omega}^2H^2/c_{\rm s0}^2$, which is equal to $2N{\tilde \omega}^2/\Omega_\bot^2$,
in equation (\ref{3.8}) weakly depends on $r$, $g$ cannot be a function of $\eta$ alone, implying that
both $g$ and $K$ also depend weakly on $r$.
Taking these weak radial dependences of $g$ and $K$ into account as a perturbation, we solve equations 
(\ref{3.8}) and (\ref{3.9}) as eigen-value problems, following Silbergleit et al. (2001).

\section{Eigen-Value Problems}

First we consider the vertical eigen-value problems.
Next, the eigen-value problems in the radial direction are considered.

\subsection{Vertical Eigen-Value Problems}

First, we solve vertical eigen-value problem, using equation (\ref{3.8}).
Since we are considering the oscillations where the vertical motions predominate over horizontal ones,
we can take $K$ to be zero in the lowest order of approximations (Silbergleit et al. 2001), and
neglect the radial dependence of ${\tilde \omega}^2H^2/c_{\rm s0}^2$, adopting its value at a radius, say $r_{\rm c}$.
Then, equation (\ref{3.8}) leads to
\begin{equation}
    (1-\eta^2)\frac{d^2g}{d\eta^2}-2N\eta\frac{dg}{d\eta}
            +2N\biggr(\frac{{\tilde\omega^2}}{\Omega_\bot^2}\biggr)_{\rm c}g=0,
\label{3.9'}
\end{equation}
where the subscript c denoted the value at the radius $r_{\rm c}$. 
At this stage, $r_{\rm c}$ is arbitrary, but later, it is determined by an eigen-value problem in the
radial direction.
The results show that $r_{\rm c}$ can be regarded as the outer capture radius of the oscillations.

Equation (\ref{3.9'}) is solved as an eigen-value problem with the boundary condition $g(\pm 1)<\infty$.
The eigenfunctions are then given by the Gegenbauer polynomials, $C_n^\lambda$ (where $n=1,2,3...$), 
i.e., 
\begin{equation}
        g(\eta)=C_n^\lambda(\eta)
\label{3.10}
\end{equation}
and the eigenvalues are
specified by
\begin{equation}
       2N\biggr(\frac{{\tilde \omega}^2}{\Omega_\bot^2}\biggr)_{\rm c}=n(n+2\lambda),
\label{3.11}
\end{equation}
where $n$ is the positive integer specifying the node number of $g$ and 
\begin{equation}
             \lambda= N-\frac{1}{2}.
\label{3.12}
\end{equation}

Equations (\ref{3.11}) and (\ref{3.12}) lead to
\begin{equation}
       {\tilde \omega}_{\rm c}^2= \Psi_n\Omega_{\bot c}^2,
\label{3.11'}
\end{equation}
where $\Psi_n$ is given by equation (\ref{1.5}), and equation (\ref{3.11'}) is nothing but 
relation (\ref{1.4}) at the capture radius.
The explicit form of $g$ is
\begin{eqnarray}
    g(\eta)\propto\left\{\begin{array}{ll} \eta & {\rm for}\quad n=1\\
                                     1-(1+2N)\eta^2 & {\rm for}\quad n=2 \\
                                     \eta-(1+2N/3)\eta^3 & {\rm for}\quad n=3.        \label{3.13}    
                    \end{array}
      \right. 
 \end{eqnarray}   

Next, a small radial variation of ${\tilde\omega}^2/\Omega_\bot^2$ is taken into account.
Then, the equation to be solved is [see equations (\ref{3.8}) and (\ref{3.9'})]
\begin{equation}
    (1-\eta^2)\frac{d^2g}{d\eta^2}-2N\eta\frac{dg}{d\eta}
            +2N\biggr[\biggr(\frac{{\tilde\omega^2}}{\Omega_\bot^2}\biggr)_{\rm c}(1+\epsilon)
                 -K(1-\eta^2)\biggr] g=0,
\label{3.14}
\end{equation}
where 
\begin{equation}
     \epsilon(r)=\frac{{\tilde \omega}^2(r)}{{\tilde\omega}^2_{\rm c}}
              \frac{\Omega^2_{\bot {\rm c}}}{\Omega^2_\bot (r)}-1.
\label{3.15}
\end{equation}
The value of $\epsilon$ is obviously zero at $r_{\rm c}$.
If we consider the radial dependences of $\Omega_\bot$ and $\Omega_{\rm K}$,
we see, after some calculation, that $\epsilon>0$ for $r<r_{\rm c}$ and $\epsilon <0$ for
$r>r_{\rm c}$, when $\omega>0$.
When $\omega<0$, however, situation is changed and $\epsilon<0$ for $r<r_{\rm c}$ and
$\epsilon>0$ for $r>r_{\rm c}$,  as far as $a_*$ is small and $\vert\omega\vert$ is not too large.

For the above perturbation method to be valid, the final results must guarrantee that $\epsilon(r)$
given by equation (\ref{3.15}) is smaller than unity in the trapped region.
The final results really show that this is the case.\footnote{
   For example, in the oscillations of $n=2$ and $n_{\rm r}=0$ (see the next subsection for the meaning of
   $n_{\rm r}$) with $\Gamma=5/3$, the range of variation of $\epsilon(r)$ in the trapped region
   is $0\sim0.12$ ($a_*=0$) and $0\sim 0.14$ ($a_*=0.2$).
   In the case of $n=3$ and $n_{\rm r}=0$ with $\Gamma=1.3$, we have $\epsilon=0\sim 0.0135$ ($a_*=0$) 
   and $0\sim0.0138$ ($a_*=0.2$).
   }
Here, we briefly mention the case of corrugation waves ($c$-mode oscillations) considered by
Silbergleit et al. (2001).
The corrugation waves are oscillations of $n=1$ and $m=1$.
In the case of $n=1$, $\epsilon(r)$ is reduced to $\epsilon(r)={\tilde\omega}^2/\Omega_\bot^2-1$, since 
$({\tilde\omega}/\Omega_\bot)_{\rm c}^2=1$ [see equations (\ref{3.11}) and (\ref{3.12})].
Hence, if $m=1$ we have $\epsilon=(\omega-\Omega)^2/\Omega_\bot^2-1$, and $\vert\epsilon(r)\vert\ll 1$ is
expected if low frequency oscillations are present.
Silbergleit et al. (2001) showed that such oscillations are really present.
This is trapping of $c$-mode oscillations.
Note, however, that in this paper we are interested in oscillations with $m=2$, not $m=1$.

Equation (\ref{3.14}) is solved by a standard perturbation method, i.e, 
the perturbed part of $g$ is expressed by a series of the Gegenbauer polynomials and 
thier coefficients as well as $K(r)$ are determined from the solvability condition of the 
inhomogeneous equation, using that 
the Gegenbauer polynomials, $C_n^\lambda(\eta)$, are orthogonal in the range of (-1,1) 
with the weight $(1-\eta^2)^{\lambda-1/2}$.
Considering this, we have, after some calculation,
\begin{equation}
      2NK(r)= n(n+2\lambda)\frac{1}{2}\epsilon(r)\chi(r),
\label{3.16}
\end{equation}
where
\begin{eqnarray}
    \chi(r)=\left\{\begin{array}{ll} 3\Gamma-1 & {\rm for}\quad n=1\\
                                     2\Gamma(5\Gamma-4)/(4\Gamma^2-7\Gamma+4) &{\rm for}\quad n=2 \\
                                     (3\Gamma-1)(7\Gamma-5)
                                       /(6\Gamma^2-9\Gamma+5) &{\rm for}\quad n=3.        \label{3.17}    
                    \end{array}
      \right. 
\end{eqnarray}  

\subsection{Radial Eigen-Value Problem}

The results in the above subsection show that the equation to be solved as the eigen-value problem
in the radial direction, i.e., equation (\ref{3.9}), is written as (Silbergleit et al. 2001),
using equation (\ref{2.3a}), (\ref{3.11}) and (\ref{3.16}), 
\begin{equation}
   \frac{1}{{\tilde \omega}}\frac{d}{dr}\biggr[\frac{{\tilde \omega}}{{\tilde\omega}^2-\kappa^2}\frac{d{\tilde h}_1}{dr}\biggr]
      +\frac{\epsilon\chi}{2c_{\rm s0}^2}{\tilde h}_1=0.
\label{3.18}
\end{equation}
Now, we introduce a new independent variable $\tau(r)$ defined by
\begin{equation}
       \tau(r)=\int_{r_{\rm i}}^r\frac{{\tilde \omega}^2(r')-\kappa^2(r')}{{\tilde \omega}(r')}dr', 
           \quad \tau_{\rm c}\equiv\tau(r_{\rm c}).
\label{3.19}
\end{equation}
Then, equation (\ref{3.18}) is written in the form:
\begin{equation}
      \frac{d^2{\tilde h_1}}{d\tau^2} +Q {\tilde h_1}=0,
\label{3.20}
\end{equation}
where
\begin{equation}
      Q(\tau)=\frac{{\tilde\omega}^2}{{\tilde\omega}^2-\kappa^2}\frac{\epsilon\chi}{2c_{{\rm s}0}^2}.
\label{3.22}
\end{equation}
Equations (\ref{3.20}) and (\ref{3.22}) show that the propagation region of oscillations is the region where
$Q>0$.
The region is that of $\epsilon>0$, which is found to be inside of $r_{\rm c}$, when $\omega>0$.
In the case of $\omega<0$, the propagation region is outside $r_{\rm c}$ and oscillations are
not trapped.

Silbergleit et al. (2001) solved equation (\ref{3.20}) by a standard WKB method with relevant boundary conditions.
The WKB approximation shows that the solution of equation (\ref{3.20}) can be represented as
\begin{equation}
    {\tilde h}_1\propto Q^{-1/4}(\tau){\rm cos}\ [\Phi(\tau)-\Phi_{\rm c}]
\label{3.21}
\end{equation}
in the whole capture region $0<\tau < \tau_{\rm c}$, except small vicinities of its boundaries
of $\tau=0$ and $\tau=\tau_{\rm c}$.
Here, $\Phi(\tau)$ is defined by
\begin{equation}
     \Phi(\tau)=\int_0^\tau Q^{1/2}(\tau')d\tau'=\int_{r_{\rm i}}^rQ^{1/2}(r')
          \frac{{\tilde \omega}^2(r')-\kappa^2(r')}{{\tilde\omega}(r')}dr',
\label{3.23}
\end{equation}
and $\Phi_{\rm c}$ is a constant to be determined by boundary conditions.
Concerning the outer boundary condition, they take into account that the capture radius, $r_{\rm c}$, 
is a turning point of equation (\ref{3.20}) since the sign of $\epsilon$ changes there.
The inner boundary condition adopted is vanishing of an arbitrary combination of ${\tilde h}_1$
and $d{\tilde h}_1/dr$ at $r_{\rm i}$.
In their treatment, $r_{\rm i}$ is taken at the marginary stable radius and $1/c_{{\rm s}0}$ is assumed 
to have a weak singularity at the radius as $1/c_{{\rm s}0}\propto (r-r_{\rm i})^{-\mu}$ with an arbitrary
parameter $\mu$.
In realistic disks, however, the inner edge is not a singularity but the temperature and density 
continue smoothly inward.
Considering this and for simplicity, we adopt $\mu=0$.
Furthermore, we consider only the case of ${\tilde h}_1=0$ or $d{\tilde h}_1/dr=0$ at the inner boundary.
Then, their results of WKB analyses show that the trapping condition is
\begin{eqnarray}
    \int_0^{\tau_{\rm c}}Q^{1/2}d\tau=\left\{\begin{array}{ll} \pi(n_r+1/4) 
                                    & {\rm for}\quad d{\tilde h}_1/dr=0 \\
                                                   \pi(n_r+3/4) 
                                    &{\rm for}\quad  {\tilde h}_1=0, 
                                                                 \label{3.24}    
                    \end{array}
      \right. 
\end{eqnarray}
where $n_r(=0,1,2,...)$ is zero or a positive integer specifying the node number of 
${\tilde h}_1$ in the radial direction.
The constant $\Phi_{\rm c}$ is also determined as
\begin{eqnarray}
    \Phi_{\rm c}=\left\{\begin{array}{ll} 0 
                                    & {\rm for}\quad d{\tilde h}_1/dr=0 \\
                                         \pi/2
                                    &{\rm for}\quad  {\tilde h}_1=0. 
                                                                 \label{3.25}    
                    \end{array}
      \right. 
\end{eqnarray}

For a given set of parameters, including spin parameter $a_*$, polytropic index $N$, 
and mass of neutron stars, $M$, any solution of equation (\ref{3.24}) specifies $r_{\rm c}$,
which gives $\omega$ of the trapped oscillation through equation (\ref{3.11}).
In other words, $\omega$ and $r_{\rm c}$ are related by equation (\ref{3.11}), i.e., 
$\omega=\omega(r_{\rm c})$ or $r_{\rm c}=r_{\rm c}(\omega)$.
Then, the trapping condition determines $r_{\rm c}$ or $\omega$ as functions of such 
parameters as $a_*$ and $N$.

\section{Numerical Results}

To obtain numerical values of the frequency, $\omega$, and the capture radius, $r_{\rm c}$, 
of trapped oscillations, we must
specify the radial distribution of acoustic speed, i.e., $c_{{\rm s}0}(r)$.
Since the final results of numerical calculations depend only weakly on the radial dependence of
$c_{\rm s0}$, we adopt the temperature distribution in the standard disk, where gas pressure dominates 
over radiation pressure and
opacity mainly comes from the free-free processes, which is (e.g., Kato et al. 2008) 
\begin{equation}
     c_{{\rm s}0}^2=1.83\times 10^{16}\Gamma (\alpha m)^{-1/5}{\dot m}^{3/5}r^{-9/10}\ {\rm cm}^2\ {\rm s}^{-2},
\label{4.1}
\end{equation}
where $\alpha$ is the conventional viscosity parameter, $m(\equiv M/M_\odot)$\footnote{
In this section and hereafter, $m$ is often used to denote $M/M_\odot$ without confusion with the
azimuthal wavenumber $m$ of oscillations.
} 
and ${\dot m}={\dot M}/{\dot M}_{\rm crit}$, ${\dot M}_{\rm crit}$ being the critical mass-flow
rate defined by 
\begin{equation}
        {\dot M}_{\rm crit}\equiv\frac{L_{\rm E}}{c^2}=1.40\times 10^{17}m\ {\rm g}\ {\rm s}^{-1},
\label{4.2}
\end{equation}
where $L_{\rm E}$ is the Eddington luminosity.
Parameters $\alpha$ and ${\dot m}$ affect on the frequencies of trapped oscillations only 
through the magnitude of $c_{{\rm s}0}$.
We adopt, throughout this paper, $\alpha=0.1$ and ${\dot m}=0.3$.
Other parameters specifying the disk-star system are $m(\equiv M/M_\odot)$ and $a_*$.
We consider the cases of $m=2.0$ and $a_*=0\sim 0.3$.

We only consider two-armed oscillations with one, two, or three node(s) in the vertical direction,
i.e., $n=1$, 2, or 3.\footnote{
In oscillations with odd number of $n$, $h_1(r,z)$ is antisymmetric with respect to the 
equatorial plane.
That is, in oscillations with $n=3$, for example, $h_1$ have one node between 
the equator and the disk surface except on the equator. 
See the functional form of $g$ given in equation (\ref{3.13}).
}
Oscillations with more nodes in the vertical direction are less interesting from
the view point of observability.
The inner boundary of oscillations is taken at the radius of $\kappa=0$, i.e.,
at the radius of the marginally stable circular orbit.
At the radius, we impose ${\tilde h}_1=0$ as the boundary condition\footnote{
  If we assume that the Lagrangian variation of pressure, i.e., $\delta p$, vanishes at
  $r=r_{\rm i}$, the condition of $h_1=0$ at $r=r_{\rm i}$ will be a better approximation 
  than $\partial h_1/\partial r=0$ at $r=r_{\rm i}$ by the following reasons.
  The Lagrangian variation of pressure, $\delta p$, can be expressed as
  $\delta p=\rho_0 h_1+\xi_r\partial p_0/\partial r+\xi_z\partial p_0/\partial z$, where 
  $\xi_r$ and $\xi_z$ are, respectively, the radial and vertical
  displacements associated with the perturbation and related to $u_r$ and $u_z$ by
  $i(\omega-m\Omega)\xi_r=u_r$ and $i(\omega-m\Omega)\xi_z=u_z$, respectively.
  If we consider that equations of motion, equations (\ref{3.2}) -- (\ref{3.4}), and that
  the pressure will not change so sharply at the $r_{\rm i}$, the major term among the above expression 
  for $\delta p$ will be $\rho_0 h_1$.
}
except in figure 5, where $d{\tilde h}_1/dr=0$ is also considered as the boundary condition 
at $r_{\rm i}$, for comparison.
The horizontal node number, $n_{\rm r}$, of oscillations we consider is mainly $n_{\rm r}=0$ and 
supplementally $n_{\rm r}=1$ and 2.

Figures 1 and 2 are the propagation diagrams for oscillations of $n=1$ and 2 (figure 1) and $n=3$
(figure 2), respectively.
Only the oscillations of $n_{\rm r}=0$ and $\Gamma=5/3$ are shown in figure 1, but
three modes of oscillations, i.e., $n_{\rm r}=0$, 1, and 2, are shown in figure 2
for the case of $\Gamma=1.25$.
In the cases of figure 1, the propagation regions of oscillations on the frequency-radius
diagram are below 
the curves of $2\Omega-\Omega_\bot$ and $2\Omega-(1+\Gamma)^{1/2}\Omega_\bot$ for $n=1$ and 
$n=2$, respectively.
The results of numerical calculations show that the oscillations of $n=1$ and $n_{\rm r}=0$ are 
trapped in the radial range shown by the upper thick horizontal line in figure 1.
The frequency $\omega$ and the capture radius $r_{\rm c}$ are, respectively, $\omega=864$Hz and 
$r_{\rm c}=3.53r_{\rm g}$.
Outside $r_{\rm c}$, the oscillation is spatially damped.
The radial range of trapped oscillations with $n=2$ and $n_{\rm r}=0$  
is shown by the lower thick horizontal line in figure 1.
The frequency and the capture radius in this case are $\omega=298$Hz and $r_{\rm c}=3.69r_{\rm g}$.

Trapped oscillations of $n=3$ have frequencies lower than those of $n=1$ and 2,
since on the propagation diagram the curve of $2\Omega-(3\Gamma)^{1/2}\Omega_\bot$ is below those of 
$2\Omega-\Omega_\bot$ and $2\Omega-(1+\Gamma)^{1/2}\Omega_\bot$ (compare figures 1 and 2).
It should be noted that the oscillation modes with $n=3$ cannot be trapped if 
$\Gamma \geq 4/3$, since in this case the propagation region is unbounded outside and
the oscillations can propagate away infinity as shown by arrow 
(see the curve of
$2\Omega-(3\Gamma)^{1/2}\Omega_\bot$ for $\Gamma=1.45$ in figure 2). 
In figure 2, the frequency and the radial extend of trapped oscillations with $n=3$
are shown for $\Gamma=1.25$ for three modes concerning the radial direction; 
the fundamental mode (i.e., $n_{\rm r}=0$) and the first two overtones
(i.e., $n_{\rm r}=1$ and 2).
The sets of frequency and capture radius for these three modes of $n_{\rm r}=0$, 1, and 2 are, respectively,
(40.8Hz, 4.30$r_{\rm g}$), (25.8Hz, 5.84$r_{\rm g}$), and (17.2Hz, 7.66$r_{\rm g}$).

Figure 3 shows the $\Gamma$-dependence of the capture radius $r_{\rm c}$.
As a typical case, the dependence is shown for oscillations with $n_{\rm r}=0$ and
some values of $n$.
It is noted that when $\Gamma$ is close to 4/3, the capture radius of oscillations with $n=3$
is far outside and their frequencies are low.
These characteristics become more prominent for oscillations with $n_{\rm r}\geq 1$, although
they are not shown in figure 3 (see figure 5).

The frequency-$\Gamma$ relations are summarized in figure 4 for a few modes of oscillations in 
two cases of $a_*=0$ and $a_*=0.1$.
Modes of oscillations adopted are $n=1$, 2, and 3.
In all cases $n_{\rm r}$ is taken to be $n_{\rm r}=0$.
As mentioned before, the oscillations with $n=3$ have low frequencies.
In order to examine characteristics of these low frequency oscillations more in detail, 
the frequency-$\Gamma$ relation in case of $n=3$ is again shown in figure 5, 
including cases where other parameter values are adopted.
That is, in addition to oscillations with $n_{\rm r}=0$, oscillations with $n_{\rm r}=1$ and 2
are considered in figure 5.
In addition, the cases where $d{\tilde h}_1/dr=0$ is adopted as the inner boundary condition 
at $r_{\rm i}$ are shown by thin curves.
In figure 6, the frequency - spin relation is shown for three modes of oscillations with
$n=1$ , 2, and 3, where $n_{\rm r}=0$ and some values of $\Gamma$ are adopted.

\begin{figure}
\begin{center}
\includegraphics[width=8cm]{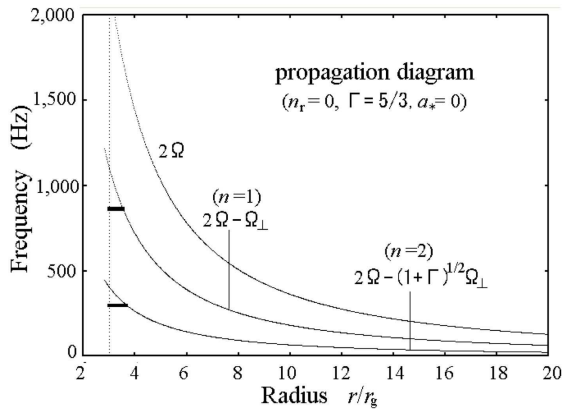}
\end{center}
\caption{
Frequency - radius plane (i.e., propagation diagram) showing the propagation region of two-armed 
nearly vertical oscillations. 
The propagation region of oscillation modes with $n=1$ is below the curve labelled by
$2\Omega-\Omega_\bot$, and the trapping of the oscillations with $n=1$ and $n_{\rm r}=0$ is shown, 
in the case of $\Gamma=5/3$, by the upper thick horizontal line 
(frequency is $\sim$ 864Hz and capture radius is 
$\sim 3.53r_{\rm g}$).
In oscillations with $n=2$, the curve specifying the boundary of the propagation region, i.e.,
$2\Omega-(1+\Gamma)^{1/2}\Omega_\bot$, depends on $\Gamma$, and the curve for $\Gamma=5/3$ is shown. 
The propagation region of oscillations with $n=2$ is below this curve in the case of $\Gamma=5/3$. 
For $\Gamma=5/3$, the trapping of the oscillations with $n=2$ and $n_{\rm r}=0$ is shown by the lower 
thick horizontal line (frequency is $\sim 298$Hz and capture radius is $\sim 3.69r_{\rm g}$).
The inner boundary condition adopted at $r_{\rm i}$ is ${\tilde h}_1=0$.
This inner boundary condition is adopted in all cases in this paper, except for in
figure 5.
The central star is assumed to have no spin.
The mass of the central star is taken to be $2M_\odot$ in all cases shown in figures in this paper.
} 
\end{figure}
\begin{figure}
\begin{center}
\includegraphics[width=8cm]{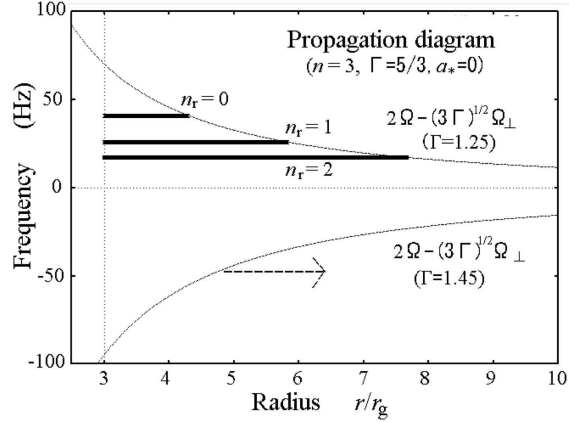}
\caption{
The same as figure 1, except that the oscillations with $n=3$ are considered here.
The propagation region of the oscillations is below the curve labelled by
$2\Omega-(3\Gamma)^{1/2}\Omega_\bot$, and the curve in the case of $\Gamma=1.25$ is shown.
Trapping of three modes of oscillations with $n_{\rm r}=0$, 1, and 2 are shown by 
three horizontal thick lines.
The sets of frequency and capture radius for these three oscillation modes are, in turn, 
(40.8Hz, 4.30$r_{\rm g}$), (25.8Hz, 5.84$r_{\rm g}$), and (17.2Hz, 7.66$r_{\rm g}$).
For the gas with $\Gamma>4/3$, the propagation region is in the outer region of disks, which is
shown by arrow, and there is no trapping.} 
\end{center}
\end{figure}
\begin{figure}
\begin{center}
\includegraphics[width=8cm]{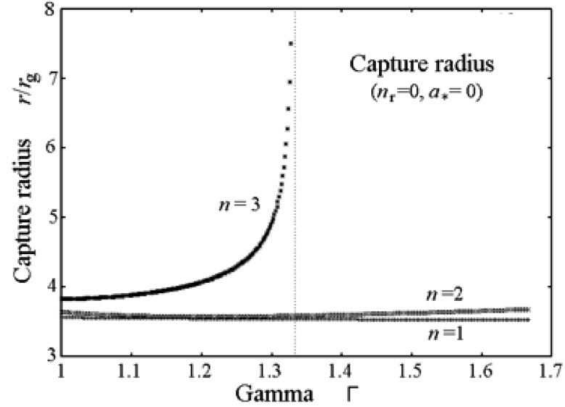}
\caption{
Capture radius, $r_{\rm c}$, as functions of $\Gamma$ for three modes of $n=1$, 2, and 3.
The radial node number $n_{\rm r}$ is taken to be zero with boundary condition ${\tilde h}_1=0$ at 
$r_{\rm i}=3r_{\rm g}$, the spin parameter being $a_*=0$. 
In oscillations with $n=3$, the trapping is absent for $\Gamma>4/3$, when $a_*=0$.   
} 
\end{center}
\end{figure}
\begin{figure}
\begin{center}
\includegraphics[width=8cm]{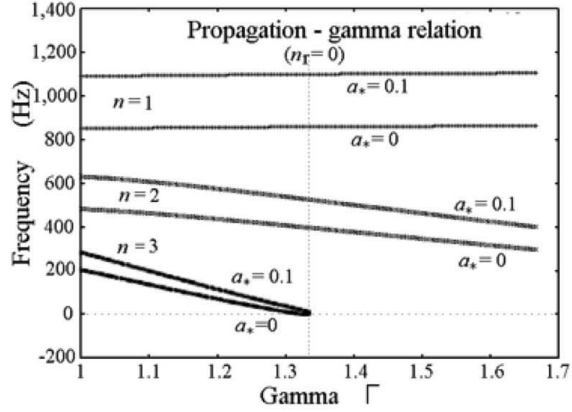}
\caption{
Frequency-$\Gamma$ relation of trapped oscillations for some values of vertical node number $n$ 
and spin parameter $a_*$.
The oscillations with no node in the radial direction ($n_{\rm r}=0$) are considered with
boundary condition of ${\tilde h}_1=0$ at $r_{\rm i}$. 
} 
\end{center}
\end{figure}
\begin{figure}
\begin{center}
\includegraphics[width=8cm]{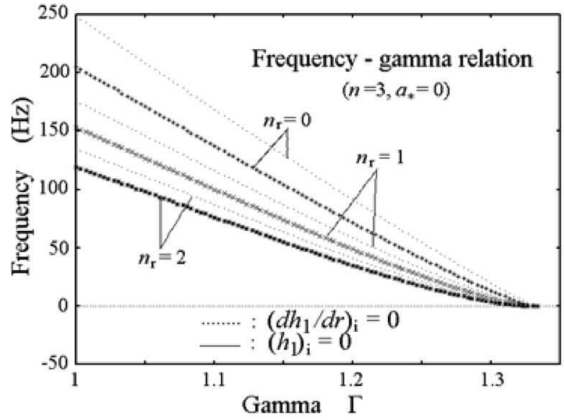}
\caption{
Frequency-$\Gamma$ relation for oscillation modes with $n=3$.
Effects of differences of radial node number $n_{\rm r}$ 
and of boundary condition on the frequency of trapped oscillations are examined. 
Two cases of boundary conditions, ${\tilde h}_1=0$ and
$d{\tilde h}_1/dr=0$ at $r_{\rm i}$, are compared for three 
modes of oscillations with radial node number $n_{\rm r}=0$, 1, and 2.
The thick curves are for the cases where the inner boundary condition is taken
as ${\tilde h}_1=0$,
while the thin curves are the cases of $d{\rm h}_1/dr=0$.
The spin parameter $a_*$ is taken to be zero. 
} 
\end{center}
\end{figure}
\begin{figure}
\begin{center}
\includegraphics[width=8cm]{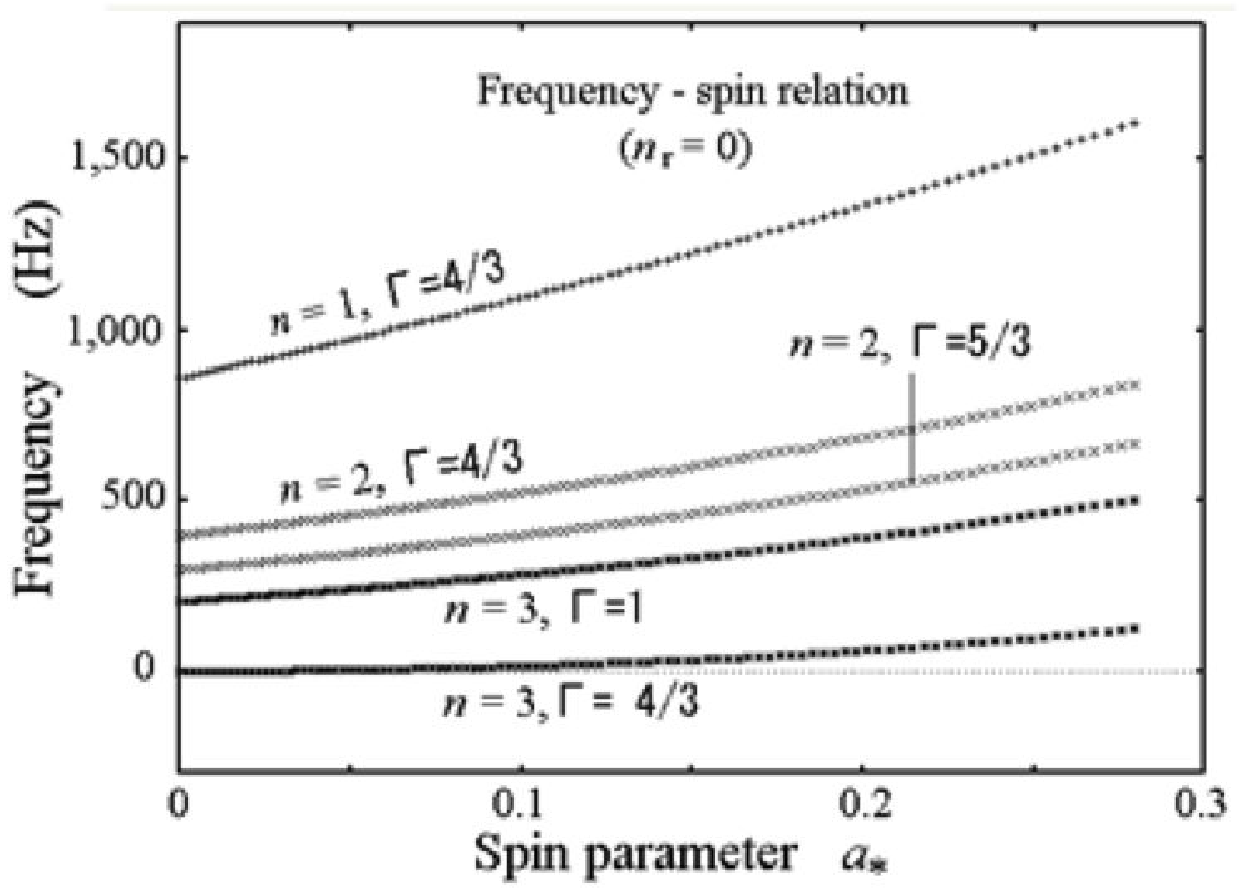}
\caption{
Frequency-spin relation for modes with $n=1$, 2, and 3.
The radial node number is taken to be $n_{\rm r}=0$.
} 
\end{center}
\end{figure}

\section{Discussion}

In this paper we have examined characteristics of two-armed ($m=2$), 
nearly vertical oscillations, assuming that the disk consists of barotropic gases with polytropic index $\Gamma$.
The parameters specifying oscillation modes are $n$ and $n_{\rm r}$, where $n(=1,2,3...)$ is 
the node number of $h_1$ in the vertical direction, and $n_{\rm r}(=0,1,2,...)$ 
is the node number of $h_1$ in the horizontal direction.
An additional important parameter is $\Gamma$, which is especially important in oscillations with $n=3$. 
Main results obtained are i) there are oscillation modes that are trapped in the inner region of disks, and 
ii) their frequencies depend on modes of oscillations ($n$, $n_{\rm r}$) and $\Gamma$, and cover a wide range of frequency.
That is, the trapped oscillations of $n=1$ and 2 have frequencies of the order of kHz QPOs, and those of 
$n=3$ are in the frequency range of the horizontal-branch QPOs (HBQPOs).

One of interesting characteristics of the oscillations is that their frequencies can change rather widely
by change of disk structure, which is distinct from the g-mode oscillations trapped around the radius of
$\kappa_{\rm max}$ (Okazaki et al. 1987), where $\kappa_{\rm max}$ is the maximum value of the 
epicyclic frequency.
That is, as shown in figure 4, the frequencies of trapped oscillations decrease by increase of $\Gamma$, 
in the $n=2$ and $n=3$ modes.
In the $n=1$ modes, however, their frequencies are insensitive to a variation of $\Gamma$ (see figure 4). 
This can be easily understood if we consider the $\Gamma$-dependence of the  
boundary curve specifying the capture radius on the propagation diagram [see figures 1 and 2, and also 
equations (\ref{1.6}) and (\ref{1.7})].

In this paper we did not quantitatively consider the effects of $c_{\rm s0}(r)$ on frequency.
An increase of $c_{\rm s0}$ without any change of other parameters leads to decrease of 
frequency of trapped oscillations.
The reason is that an increase of $c_{\rm s0}$ decreases $Q$.
Hence, to satisfy the trapping condition (\ref{3.24}), an increase of $r_{\rm c}$ is necessary,
which leads to decrease of frequency (see figures 1 and 2).
In table 1, the effects of changes of various parameter values on frequencies of trapped oscillations are summarized.

\begin{table}
\caption{table 1.}
\begin{center}
\begin{tabular}{llll}


\hline\hline
    parameters  &     & frequency  \\  \hline
    parameters of oscillations & increase of $n$ (except for $n=1$)      &   decrease \\
                               & increase of $n_{\rm r}$                 &   decrease \\
    \hline
    disk parameters            & increase of $\Gamma$ (except for $n=1$) &   decrease \\
                               & increase of $c_{\rm s0}$     & decrease     \\
    \hline
    parameters of stars        & increase of $m\equiv (M/M_\odot)$  & decrease     \\
                               & increase of $a_*$            & increase     \\ 
    \hline
\end{tabular}
\end{center}
\end{table}

One of important problems remained is whether the nearly vertical oscillations considered in this paper 
can be really excited on disks.
Two possibilities will be conceivable.
One is the excitation by the process of viscous overstability of oscillations (Kato 1978).
This, however, might be inefficient to some types of oscillations, especially to oscillations 
with large $n$.
Another and more promising process is the stochastic excitation of oscillations by turbulence, developed first by 
Goldreich and Keely (1977a,b).
This is known as the excitaion process of solar and stellar non-radial oscillations.
This process is better than the former in the sense that it will be able to excite many types of oscillations,
without no particular selection concerning the forms of eigenfunctions.

So far, we did not discuss the effects of corotation resonance on the present trapped oscillations.
It is known that non-axisymmetric g-mode oscillations are generally damped by corotation
resonance (Kato 2003; Li et al. 2003; Latter \& Balbus 2009; see also Silbergleit \& Wagoner 2008).
The c-mode oscillations, which are non-axisymmetric, are also damped by the resonance
(Tsang \& Lai 2009).
These results are related to the fact that these oscillations have node(s) in the vertical direction.
The vertical oscillations considered in this paper have also node(s) in the vertical direction.
The oscillations, however, have an important difference from g- and c-mode oscillations.
That is, in the case of two-armed nearly vertical oscillations, the radius of coratation resonance, 
i.e., the radius where $\omega=2\Omega$ is realized, is far outside the propagation region
(see the curve of $2\Omega$ in figure 1; the curve of $2\Omega$ is not shown in figure 2, 
since it is outside the diagram).
In the evanescent region the wave amplitude is spatially damed exponentially.
Hence, the effects of corotation damping are negligible in the present oscillations. 

It will be important to note here that the effects of the general relativity are not essential in the
trapping of the present vertical oscillations, except for the modes with $n=1$.
The propagation region of the vertical oscillations is specified by inequality (\ref{1.8}).
In deriving inequality (\ref{1.8}), the fact that $\Psi_n\Omega^2_\bot$ is larger than 
$\kappa^2$ is adopted.
In the case of oscillation modes of $n=1$, the general relativity is necessary to guarantee this, since
$\Psi_n=1$.
In oscillations with $n\geq 2$, however, $\Psi_n\Omega^2_\bot>\kappa^2$ is guarantees even in 
Newtonian disks.
Other important ingredients for presence of trapped oscillations are the presence 
of inner edge of disks where waves are reflected back and
a monotonical decrease of angular velocity of rotation, $\Omega(r)$, outwards.
This point is different from trapping of g-mode oscillations, since
in trapping of g-mode oscillations, the general relativity is essential. 
If we want to describe the QPOs in disks extending from black-hole or neutron star systems to dwarf-novae
systems by a common mechanism, the nearly vertical oscillations considered here will be one of good candidates, since
in the disks of dwarf-novae the effects of the general relativity are minor.

We have shown that there are various modes of nearly vertical oscillations trapped in the inner region of disks.
From the observational points of view, however, the oscillations with $n=1$ and $n=2$ may not be
so interesting, since their trapped regions are too narrow as shown in figures 1 and 3, and may not be 
observed with large amplitudes, although the trapped regions are close to the inner edge of disks.
Compared with them, the oscillations with $n=3$ will be of interest, since their trapped region
is wide as shown in figures 2 and 3.
In this sense, the $n=3$ modes will be one of possible candidates of low frequency QPOs
such as horizontal-branch QPOs (HB QPOs).
Further discussion on this direction will be worthwhile.
Related to this, there are some issues to be noted here.

First, we have found that low frequency oscillations with $n=3$ are present only in disks with
$\Gamma<4/3$.
Hence, one may imagine that such oscillations cannot be expected in neutron-star disks, since
the standard disk model shows that the main part of the disks is gas-pressure-dominated.
We should notice, however, that in disks with high accretion rate (accretion rate is close to or higher than
the critical accretion rate defined by the Eddington luminosity),
a radiation-pressure-dominated region appears in the innermost part of disks.
In particular, in slim disks the whole region is radiation-pressure-dominated.
Furthermore, we should notice that it is unnecessary to regard the index $\Gamma$ as the same as the ratio
of the specific heats, $\gamma$, of the disk gas.
In practice, we can expect $\Gamma<\gamma$ by the following reasons.
Due to radiation from a hot corona, disks have tendency to approach isothermal disks in the vertical direction.
Furthermore, in low frequency oscillations, radiative heat transport tends to make the oscillations isothermal,
although as another effect it may dampen oscillations.

Second, the oscillations with $n=3$ are trapped in a rather wide region,, i.e., the radial wavelength is long.
Hence, it will be necessary to take into account some terms neglected in deriving equation (\ref{3.7-}),
including the terms of relativistic corrections, when we want to do more quantitative estimate of
oscillation frequencies.

Finally, we should mention a difference between the low frequency oscillations with $n=3$ discussed  
in this paper and the low-frequency one-armed corrugation waves (Kato 1989, Silbergleit et al. 2001).
The latter oscillations are a kind of warps or tilts, and roughly an incompressible deformation of disks,
while the former are compressible oscillations.

\bigskip
The author thanks the referee for valuable comments  with careful reading of the original version.

\bigskip
\leftskip=20pt
\parindent=-20pt
\par
{\bf References}
\par
Ferreira, B.T. \& Ogilvie, G.I. 2008, MNRAS, 386, 2297 \par
Goldreich, P. \& Keely, D.A. 1977a, ApJ, 211, 934 \par
Goldreich, P. \& Keely, D.A. 1977b, ApJ, 212, 243 \par
Henisey, K.B., Blaes, O.M., Fragile, P.C., \& Ferreira, B.T. 2009, arXiv:0910.1882 (to be published in ApJ) \par
Kato, S. 1978, MNRAS, 185, 629 \par
Kato, S. 2001, PASJ, 53, 1\par 
Kato, S. 2003, PASJ, 55, 257 \par
Kato, S. 2004, PASJ, 56, 905\par
Kato, S. 2005, PASJ, 57, 699 \par
Kato, S. 2008a, PASJ, 60, 111 \par
Kato, S. 2008b, PASJ, 60, 1387 \par
Kato, S., Fukue, J. 2006, PASJ, 58, 909\par
Kato, S., Fukue, J., \& Mineshige, S. 2008, Black-Hole Accretion Disks --- Towards a New paradigm --- 
  (Kyoto: Kyoto University Press)\par
Latter, H.N. \& Balbus, S.A. 2009, MNRAS, 399, 1058 \par 
Li, L.-X., Goodman, J., Narayan, R. 2003, ApJ, 593,980 \par
Nowak, M.A \& Wagoner, R.V. 1992, ApJ, 393, 697 \par
Okazaki, A.T., Kato, S., \& Fukue, J. 1987, PASJ, 39, 457\par
Oktariani, F., Okazaki, A.T. \& Kato, S. 2010, submitted to PASJ \par
Perez, C.A., Silbergleit, A.S., Wagoner, R.V., \& Lehr, D.E. 1997, ApJ, 476, 589 \par
Silbergleit, A.S., Wagoner, R., \& Ortega-Rodriguez, M. 2001, ApJ, 548, 335 \par
Silbergleit, A.S. \& Wagoner, R. ApJ, 680, 1319 \par
Tsang, D. \& Lai, D. 2009, MNRAS, 393, 992 \par 
\bigskip\bigskip

\end{document}